\documentclass{article}
\usepackage{spconf,amsmath,graphicx}
\usepackage{multirow}
\usepackage{lscape,array}
\usepackage{booktabs}
\usepackage{pifont}
\usepackage{ragged2e}
\usepackage{amssymb}
\usepackage{enumitem}

\title{Target Speech Extraction with pre-trained \\self-supervised learning models}
\name{Junyi Peng$^{1}$, Marc Delcroix$^{2}$, Tsubasa Ochiai$^{2}$, Old\v{r}ich Plchot$^{1}$, Shoko Araki$^{2}$, Jan \v{C}ernocký$^{1}$}

\address{
$^1$Brno University of Technology, Faculty of Information Technology, Speech@FIT, Czechia \\
$^2$NTT Corporation, Japan\\
}
\ninept
\newcolumntype{L}[1]{>{\raggedright\let\newline\\\arraybackslash\hspace{0pt}}m{#1}}
\newcolumntype{C}[1]{>{\centering\let\newline\\\arraybackslash\hspace{0pt}}m{#1}}
\newcolumntype{R}[1]{>{\raggedleft\let\newline\\\arraybackslash\hspace{0pt}}m{#1}}
\begin{document}
%\ninept
%
\maketitle
\begin{abstract}
Pre-trained self-supervised learning (SSL) models have achieved remarkable success in various speech tasks. However, their potential in target speech extraction (TSE) has not been fully exploited. TSE aims to extract the speech of a target speaker in a mixture guided by enrollment utterances. We exploit pre-trained SSL models for two purposes within a TSE framework, i.e., to process the input mixture and to derive speaker embeddings from the enrollment. In this paper, we focus on how to effectively use SSL models for TSE. We first introduce a novel TSE downstream task following the SUPERB principles. This simple experiment shows the potential of SSL models for TSE, but extraction performance remains far behind the state-of-the-art. We then extend a powerful TSE architecture by incorporating two SSL-based modules: an Adaptive Input Enhancer (AIE) and a speaker encoder. Specifically, the proposed AIE utilizes intermediate representations from the CNN encoder by adjusting the time resolution of CNN encoder and transformer blocks through progressive upsampling, capturing both fine-grained and hierarchical features. Our method outperforms current TSE systems achieving a SI-SDR improvement of 14.0 dB on LibriMix. Moreover, we can further improve performance by 0.7 dB by fine-tuning the whole model including the SSL model parameters.
%Specifically, the proposed AIE processes the mixture and employs progressive upsampling to adjust the time resolution of the layers of the CNN Encoder and Transformer blocks of the SSL model, thus capturing richer hierarchical features.
\end{abstract}
\begin{keywords}
Target speech extraction, pre-trained models, self-supervised learning, feature aggregation
\end{keywords}
\section{Introduction}
\label{sec:intro}
Over the past several years, transformer models trained with self-supervised learning (SSL) \cite{hsu2021hubert, baevski2020wav2vec} have shown great success in various speech tasks, such as automatic speech recognition (ASR) \cite{arunkumar2022joint}, speaker verification (SV) \cite{vaessen2022fine, chen2022large}, and speech enhancement (SE) \cite{song2023exploring, hung2022boosting}. This effectiveness is attributed to the models' ability to learn overcomplete and general-purpose features when pre-trained on large-scale datasets, thereby ensuring robust performance and generalization, even under data-limited conditions \cite{peng2023improving}. 

% Recently, pre-trained SSL models have also been used for speech enhancement (SE), such as noise reduction \cite{song2023exploring, hung2022boosting} or speech separation \cite{chen2023speech} with relative success.
Despite their strong performance in several downstream tasks, there are only a few studies investigating the use of SSL representations for target speech extraction (TSE) \cite{liu2023quantitative}, a task aiming at estimating the speech of a target speaker from a multi-talker mixture\cite{vzmolikova2023TSEoverview}.
%However, the incorporation of SSL representations in Target Speech Extraction (TSE) remains an underexplored area \cite{}.

TSE approaches usually use an extractor module consisting of a neural network, which inputs a speech mixture and estimates the target speech by exploiting a speaker embedding derived from an enrollment of the target speaker to identify him/her in the mixture. Consequently, the TSE approaches are related to separation and SV, making it an interesting use case for pre-trained SSL models. In this paper, we explore using pre-trained SSL models for TSE. 

%One way to design a SSL-based TSE system is using a simple downstream model, such as a stack of bidirectional long short-term memory (BLSTM) following principles of Speech processing Universal PERformance Benchmark (SUPERB) \cite{yang2021superb, tsai2022superb}. 
Leveraging the principles of Speech processing Universal PERformance Benchmark (SUPERB) \cite{yang2021superb, tsai2022superb}, we propose an SSL-based TSE system with a straightforward downstream model, such as a stack of bidirectional long short-term memory (BLSTM) for the extractor.
The SSL model plays a dual role: first, extracting features from the input mixture and, second, obtaining speaker embeddings from the enrollment utterance. 
%[Here you can briefly describe the SUPERB-based TSE system, including the extraction and speaker encoder part. The point is that the SSL is used twice].
We show experimentally that such a system can be used to design a TSE system. However, as for other SE tasks, the performance is far behind the state-of-the-art. 

One potential reason for this performance gap could be the large strides used in SSL models, which typically operate with a stride of about 20ms, yielding only 50 frames per second. This is in stark contrast to widely-used SE models, such TasNet \cite{luo2018tasnet}, which utilizes smaller strides of 1.25ms (i.e. 800 frames per second), thus have better temporal resolution that might be crucial for optimal performance in SE tasks. 
Moreover, it is worth noting that SSL models like WavLM \cite{hsu2021hubert, song2023exploring} typically consist of two main components: a CNN Encoder and a series of Transformer blocks. Recent studies \cite{hung2022boosting, huang2022investigating, chen2022wavlm} have indicated that lower layers of the SSL models, especially the outputs of the CNN encoder that serve as the input to the Transformer blocks, are more relevant for SE tasks. Despite this, most SE models \cite{huang2023adapting} typically focus on the representations obtained from the Transformer layers, neglecting the outputs of the intermediate CNN Encoder layers, thus failing to take advantage of the hierarchical representations acquired by models pre-trained on large-scale datasets. 

%Additionally, it is notable that the incorporation of SSL representations in Target Speech Extraction (TSE) remains an underexplored area. In detail, limited work exists where the SSL model serves dual roles as both the speaker encoder and extractor, presenting significant room for further research.

To tackle the aforementioned challenges, this paper presents a systematic approach that leverages multi-scale representations from SSL models for TSE tasks. 
%Specifically, we first follow the SUPERB principles to introduce a novel downstream task called TSE, aimed at providing a comprehensive evaluation of existing SSL models. Subsequently, for integration with current TSE systems, 
We construct two modules based on a frozen pre-trained Transformer model named the \emph{speaker encoder} (SpkEnc) and the \emph{Adaptive Input Enhancer} (AIE). The SpkEnc module computes target speaker embeddings from the enrollment by following prior works on using SSL models for SV \cite{peng2023attention}. The AIE module 
is designed to extract features from the mixture. In particular, it adjusts the time resolutions across intermediate layers of the CNN Encoders and the transformer of the SSL model, thereby allowing the exploitation of multi-scale feature representations.
%These modules, named the speaker encoder and the Adaptive Input Enhancer (AIE), are designed to extract target speaker embeddings and enrich features for extraction, respectively. 
Finally, the entire SSL model is jointly fine-tuned with the TSE system to enhance performance further. Overall, our contributions are as follows:
\begin{itemize}[itemsep=2pt,topsep=0pt,parsep=0pt]
%\textbf{TSE downstream tasks.} A
\item  We propose a new TSE downstream task, developed in line with SUPERB principles, which goes beyond single-objective downstream tasks 
%by focusing on generative and discriminative capabilities of pre-trained models.
by emphasizing the multi-faceted capabilities of pre-trained models, such as SE feature extraction and speaker encoding.
\item %\textbf{Integration with TSE systems.} Leveraging a shared frozen pre-trained SSL model, 
We introduce two sub-modules, i.e., SpkEnc and AIE, designed to function as plug-in units, enabling flexible integration of the pre-trained SSL model into a powerful TSE architecture. %In particular, the AIE module is designed to adjust the time resolutions across intermediate layers of the CNN Encoders and the transformer, thereby allowing the exploitation of multi-scale feature representations.
%\item  \textbf{Incorporation of multi-scale information.} We explore U-Net \cite{ronneberger2015u} and Feature Pyramid Network (FPN) \cite{lin2017feature} structure in the AIE module to adjust the time resolutions across intermediate layers of the CNN Encoders, thereby enhancing the modeling of hierarchical multi-scale representation.
\item We conduct a comprehensive study on the Libri2mix dataset and demonstrate that exploiting pre-trained SSL representation can boost the performance of a powerful TSE system, %. Our proposed TSE system significantly outperforms 
outperforming prior systems such as SpEx+ \cite{ge2020spex} and TD-SpeakerBeam \cite{delcroix2020improving}.
% on the TSE performance of pre-trained models under various learning conditions, including partial freezing or full fine-tuning of the Speaker Encoder and AIE modules.
% \item Through extensive experiments on the Libri2mix dataset, we demonstrate that our approach significantly outperforms current TSE systems such as SpEx+ \cite{ge2020spex} and TD-SpeakerBeam \cite{delcroix2020improving} systems.
\end{itemize}

\section{Prior works}
%\subsection{Exploiting pre-trained SSL models for SE}
Pre-trained SSL models have been used for SE tasks such as denoising and speech separation. 
In \cite{huang2022investigating}, SSL model representations are used to estimate the time-frequency mask for the Short-Time Fourier Transform (STFT) of the input signal, refined by BLSTM layers. SSL-based approach demonstrated superior performance compared to FBANK-based methods on LibriMix \cite{cosentino2020librimix} and Voicebank-DEMAND \cite{veaux2013voice}. To further improve SE performance, various strategies have been explored, such as the fusion of SSL and STFT features \cite{hung2022boosting}, and the introduction of a regression-based training objective \cite{valentini2016investigating}. However, unlike our proposal, they do not exploit the features from the CNN layers, although they are probably the most relevant for SE tasks.

The only prior work using SSL for TSE is \cite{liu2023quantitative}, where a pre-trained SSL model is only employed to derive the speaker embeddings from the enrollment, resulting in a limited improvement (0.3 dB) compared to using FBANK features.
%The authors XXX. % However, the application of SSL models in TSE remains a less-studied area that we aim to explore in the following sections.

%\subsection{SSL models for Target Speaker-ASR}
SSL models have also been used for target speaker-ASR, which focuses on transcribing a specific speaker from a segmented utterance containing multi-talker speech using enrollment speech for that speaker. In \cite{huang2023adapting}, a speaker embedding is prepended to the input features of the SSL model's Transformer blocks. The entire model is then fine-tuned using a CTC loss. This approach differs from TSE as it outputs a character sequence instead of the target speech signal.
%[Also mention the TS-SSL models for TS-ASR, which shares similarity with TSE, but performs ASR directly instead of estimating the speech signals]

\section{Conventional neural TSE}
Let us first describe the overall architecture of a typical TSE system \cite{vzmolikova2023TSEoverview}. A neural TSE system consists of four main blocks: the encoder, decoder, SpkEnc, and extractor. The encoder transforms the input speech mixture $\mathbf{y}$ into higher-dimensional features $\mathbf{Z}_y$ as $\mathbf{Z}_y=\text{Encoder}(\mathbf{y})$, which could be either spectral features obtained via STFT or learned features derived from 1D convolutional layers operating on the raw waveform. SpkEnc is responsible for extracting speaker embedding $\mathbf{e}$ that captures the voice characteristics of the target speaker, usually derived from an enrollment speech $\mathbf{c}$, as $\mathbf{e}=\text{SpkEnc}(\mathbf{c})$. The extractor estimates the target speech from the mixture in the feature domain $\mathbf{Z}_y$, where $\mathbf{Z}_s=\text{Extractor}(\mathbf{Z}_y,\mathbf{e})$, given the target speaker embeddings $\mathbf{e}$. Finally, the decoder transforms the processed feature $\mathbf{Z}_s$ into the estimated  target speech $\hat{\mathbf{x}}$, where $\hat{\mathbf{x}}=\text{Decoder}(\mathbf{Z}_s)$. 
%You need an explanation of a conventional TSE system so that the reader can understand your contribution. This can be simple but should include a brief description of encoder/decoder, extraction with mask, speaker embedding extraction and conditioning.

\section{Exploiting pre-trained SSL models for TSE}
Many SSL models including WavLM \cite{chen2022wavlm}, Hubert \cite{hsu2021hubert}, and \linebreak wav2vec2.0 \cite{baevski2020wav2vec} consist of CNN and Transformer blocks producing the intermediate outputs denoted $\mathbf{H}^{\text{cnn}}_{j}$ and $\mathbf{H}^{\text{trf}}_{i}$, with $j\in\{1, \ldots, J\}$ and $i\in\{1, \ldots, N\}$ indicating the index of CNN and Transformer blocks, respectively, where $J$ and $N$ denote the total number of CNN and Transformer blocks in the models.
We can exploit the feature representation obtained by these models as input features for the extractor and for the SpkEnc. 

First, we propose a simple TSE downstream model following the style of the SUPERB evaluation to carry out preliminary experiments. We then discuss how to exploit SSL features within a more powerful TSE architecture.

%Firstly, we introduce a novel downstream model devised explicitly for TSE, which serves as the first attempt to use SSL models in dual roles, i.e., both as a generator of speech signals and as a feature extractor for speaker-related characteristics. Subsequently, our focus shifts to integrating SSL models within an existing TSE architecture, as depicted in Fig \ref{fig:sys}. Given the dependency on the architecture of upstream SSL models, a brief description is in order. Hubert variants including WavLM Base, WavLm Base Plus, and wav2vec2.0 consist of CNN and Transformer blocks producing the intermediate outputs denoted $\mathbf{H}_{\text{cnn}}^{j}$ and $\mathbf{H}_{\text{trf}}^{i}$, with $j$ and $i$ indicating the index of CNN and Transformer blocks, respectively.  
\begin{figure}[t]
    \centering
    \includegraphics[width=0.99\linewidth]{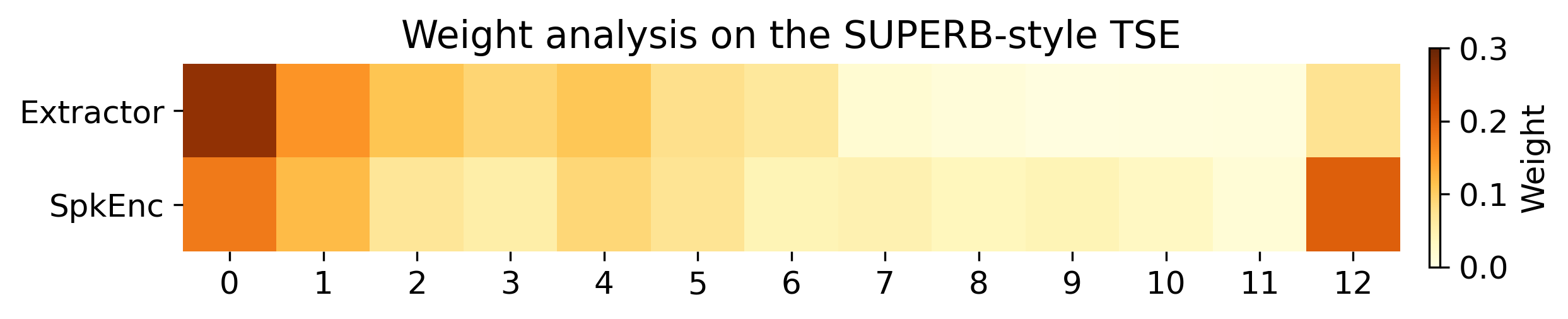}
        \vspace{-0.2cm}
    \caption{Layer-wise weights of speaker encoder (SpkEnc) and extractor, using the BLSTM-based TSE downstream model, and WavLM Base Plus pretrained SSL model. Note that $0$-th Transformer layer denotes the output of the CNN encoder, which is also the input of the 1st Transformer layer.}
    \label{fig:2}
        \vspace{-0.4cm}
\end{figure}
\begin{figure*}[t]
    \centering
    \includegraphics[width=0.9\linewidth]{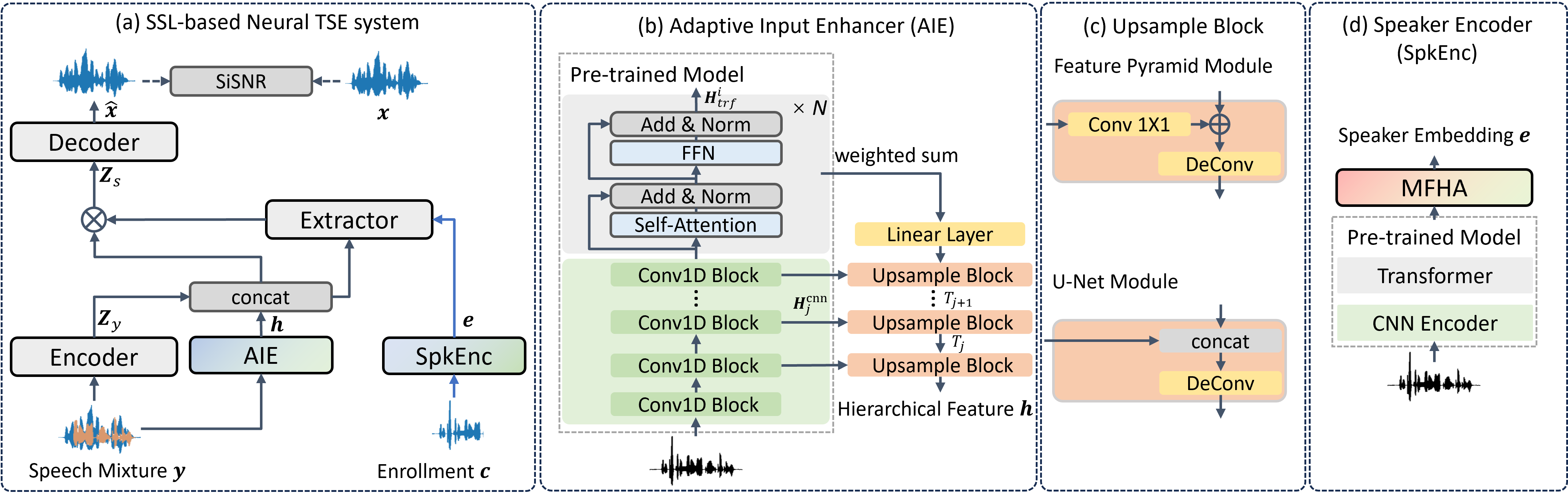}
        \vspace{-0.3cm}
    \caption{(a) Overview of the proposed SSL model-based TSE system, and the details of (b) AIE module, (c) upsample blocks, and (d) SSL-based SpkEnc.}
    \label{fig:sys}
    \vspace{-0.6cm}
\end{figure*}

\subsection{SUPERB-style downstream TSE model}
\label{sec:SUPERB}
In line with SUPERB's methodology, we construct two lightweight modules that utilize two different set of weights to perform a weighted sum of SSL features as $\sum w_{i}^{\nu}\mathbf{H}^{\text{trf}}_{i}$, where $\sum w_i^{\nu}=1$, where $w_i^{\nu}$ are the weights for the extractor or SpkEnc (i.e., $\nu=\{\text{SpkEnc}, \text{extractor}  \}$) for layer $i$ of the Transformer. The SpkEnc computes the target speaker embedding by averaging these weighted features over frames followed by a linear layer. The extractor comprises a three-layer BLSTM model that predicts a time-frequency mask.
It accepts the SSL features as input. The processing is conditioned on the target speaker by multiplying the speaker embedding with the output of the first BLSTM \cite{vzmolikova2019speakerbeam}. We then multiply the STFT coefficients of the speech mixture with the estimated mask and apply inverse STFT to produce the waveform of the extracted speech. 
During the training, the SSL model is kept frozen while the sub-modules, including the weights $w_i^{\nu}$ are learnable. 
We provide experiments with such a model in Section \ref{sec:superb_exp}. 

Despite its simplicity, it succeeds in performing TSE, but the performance remains far behind current TSE systems trained from scratch (e.g., TD-SpeakerBeam \cite{delcroix2020improving}). Figure \ref{fig:2} plots the weights $w_i^{\nu}$ obtained with such a system. This reveals that lower layers of the Transformer blocks, particularly layer 0, corresponding to the output of the CNN encoder, are more critical for the extractor.
% SHALLOWER ——> ??

%SSL features are used in two ways, as feature extraction for the mixture and to compute the speaker embeddings.
%Here you first describe how each block of conventional TSE is implemented. It works in the frequency domain, estimate mask with BLSTM, condition by multiplication of the embedding after first BLSTM...
%[You can reuse some parts of your explanation below]

%This is a simple way to exploit SSL model, but it may not reach high performance.

\subsection{TD-SpeakerBeam extension with pre-trained SSL models}
%Briefly describe TD-SpeakerBeam [could be explained in the previous section too].
We then explore effectively exploiting SSL representation within a powerful TSE system, such as TD-SpeakerBeam~\cite{delcroix2020improving}. TD-SpeakerBeam uses 1-D convolutions for the encoder and decoder and temporal convolutional network (TCN) blocks for the extractor and SpkEnc. Note that the time resolution of TD-SpeakerBeam is typically much higher than most layers of an SSL model.

Building upon the insights from the SSL weight visualization of the SUPERB-style TSE model in Fig. \ref{fig:2},
%Building upon the insights presented in Fig \ref{fig:2}, 
%which suggests that the shallower layers of the Transformer blocks are more impactful for Extractor that require fine-grained information. W
we investigate the use of the CNN encoder module positioned before the Transformer blocks of the SSL model. We introduce an AIE module that integrates the output of the different layers of the SSL model. In particular, AIE performs progressive upsampling to adjust the time resolutions of the different layers of the CNN encoder and the Transformer, capturing multi-scale and fine-grained information from the SSL module.
%Due to the varying time resolutions across intermediate layers of the CNN Encoder, we adopt a progressive upsampling approach to incorporate both fine-grained and multi-scale features effectively. We further propose an Adaptive Input Enhancer (AIE) to further leverage these diverse representations from both the Transformer blocks and CNN Encoder layers, capturing multi-scale and fine-grained information. 
The output of the AIE module, $\mathbf{h}$, has the same time resolution as the output features of the encoder, $\mathbf{Z}_y$. Consequently, these two streams can be concatenated along the feature dimension and fed into the Extractor model as $\mathbf{Z}_s=\text{Extractor}(\text{concat}(\mathbf{Z}_y,\mathbf{h}))$, as shown in Fig. \ref{fig:sys}-(a)

\subsubsection{Adaptive input enhancer}
The proposed AIE leverages a series of one-dimensional deconvolutional blocks formulated as $\mathbf{T}_j=\text{Upsample}(\mathbf{T}_{j+1}, \mathbf{H}^{\text{cnn}}_{j})$, where $\mathbf{T}_j$ is the output of $j$-th upsample block and has the same time resolution as $\mathbf{H}^{\text{cnn}}_{j}$, as shown in Fig. \ref{fig:sys}-(b). The output of AIE is the hierarchical feature obtained as $\mathbf{h}=\mathbf{T}_2$, since the second CNN layer has the same time resolution as $\mathbf{Z}_y$ for the typical setting of TD-SpeakerBeam. Note that when considering features from both the Transformer and the CNN, the topmost layer of the AIE is initialized as $\mathbf{T}_{top}=\text{Linear}(\sum_{i=1}^{N} w^{\text{AIE}}_i \mathbf{H}^{\text{trf}}_i)$, where $w^{\text{AIE}}_i$ are learnable scalar weights summing to one. 
%Otherwise, it defaults to the output from the final layer of the CNN.

We experiment with two variants to implement the upsampling operation 
%borrowed from the  FPN\cite{lin2017feature} and U-Net\cite{ronneberger2015u} architectures 
as shown in Fig. \ref{fig:sys}-(c). %\textbf{Unet-Style Module}: This sub-module uses stacked deconvolution blocks to merge features from corresponding encoding layers in the CNN Encoder. It aims to incorporate the contextual information from the encoding layers while preserving fine-grained details by skip connections between layers.

\textbf{Feature Pyramid Module}: The upsampling operation is borrowed from the FPN architecture \cite{lin2017feature}. It is implemented as follows: $\mathbf{T}_j= \text{DeConv}(\text{Conv}(\mathbf{H}^{\text{cnn}}_{j}) + \mathbf{T}_{j+1})$, where $\text{DeConv}(\cdot)$ is a deconvolution operation that perform upsampling and $\text{Conv}(\cdot)$ is a convolution operation used to transform the output of the $j$-th CNN block. 
%This sub-module employs a top-down architecture to extract multi-scale representations from the CNN Encoder layers. This setup enables the AIE to yield hierarchical features to boost the Extractor's performance in TSE tasks. 

\textbf{Unet-Style Module}: This upsampling alternative follows the U-Net architecture \cite{ronneberger2015u} and is implemented as: $\mathbf{T}_j = \text{DeConv}(\text{Concat}(\\\mathbf{H}^{\text{cnn}}_{j}, \mathbf{T}_{j+1}))$, where the concatenation is performed along the channel dimension.

The implementation of the upsampling slightly differs between these two configurations, but both approaches perform top-down processing to extract multi-scale representations from the CNN Encoder layers. We expect that such a hierarchical upsampling process can capture rich information from the pre-trained SSL model, yielding hierarchical features, $\mathbf{h}$, that can complement the features obtained from the TSE encoder, $\mathbf{Z}_y$. %to boost the Extractor's performance in TSE tasks. 
%This sub-module uses stacked deconvolution blocks to merge features from corresponding encoding layers in the CNN Encoder. It aims to incorporate the contextual information from the encoding layers while preserving fine-grained details by skip connections between layers.

%When considering features from both the Transformer and the CNN, the topmost layer of the AIE is initialized as $T_{top}=\text{Linear}(w\mathbf{H}_{trf})$. Otherwise, it defaults to the output from the final layer of the CNN. 

\subsubsection{Speaker encoder based on pre-trained model}
\label{sec:spkenc}
Compared to SUPERB architecture, which simply adopts the average operation, we employ an advanced attentive pooling, named multi-head factorized attentive pooling (MHFA) \cite{peng2023attention}, to enhance the quality of learned speaker representation, as shown in Fig. \ref{fig:sys}-(d). This SSL-MHFA employs two sets of normalized layer-wise weights to generate attention maps and compressed features, which are expected to encode speaker-discriminative information and phonetic information respectively.  Then, the speaker embedding is formed by aggregating over frames and projecting the vector to a lower-dimensional space using a linear layer. In this way, each attention head is expected to aggregates information from a specific set of phonetic units, which leads to a robust speaker embedding.

\section{Experiments}
\vspace{-2mm}
\subsection{Experiment Setup}
\vspace{-1.5mm}
\textbf{Data-sets:}
We conduct experiments using the Libri2Mix dataset, consisting of simulated mixtures of two speakers\cite{cosentino2020librimix}. Following the data preparation in TD-SpeakerBeam\footnote{https://github.com/BUTSpeechFIT/speakerbeam} with 16kHz sampling rate, dataset is partitioned into three subsets: train-100, dev, and test.
%The train-100 is employed for model training and comprises a total of 58 hours of utterances from 291 speakers. Both the dev and test subsets contain 40 unseen speakers each with 11 hours. 
We choose the model with the best SI-SNRi performance on the dev.

\noindent\textbf{Implementation details:} For \emph{SUPERB-style setup}, we use a BLSTM-based TSE. The window size and the number of FFT points are set to 1024, the dimension of BLSTM is 512, and so is the speaker embedding. As a baseline, we use the same architecture with STFT or FBANK features. 
%In that case, SpkEnc is implemented as Sec \ref{sec:SUPERB}.

For the \emph{TD-SpeakerBeam setup}, we use an SSL-based speaker encoder using MHFA with a total of 8 heads as in \cite{peng2023attention}. The speaker embedding is of dimension 256. 
%For AIE, we select the output from the deconvolution block corresponding to the second convolutional layer in CNN Encoder as the hierarchical representation, due to its temporal resolution matching that of the TSE system. 
For TD-SpeakerBeam baseline, the configuration follows the specifications in \cite{delcroix2020improving}. We use Adam as the optimizer with an initial learning rate set to $10^{-3}$. When the pre-trained model is unfrozen for joint optimization with the TSE system, the learning rate is set to $2\times10^{-5}$. When using TD-SpeakerBeam as downstream model, we use the WavLM Base Plus\footnote{https://huggingface.co/microsoft/wavlm-base-plus}.

\noindent\textbf{Performance Metrics:} We measure performance in terms of source-to-distortion Ratio (SDR), scale-invariant SDR improvement (SI-SDRi), perceptual evaluation of speech quality (PESQ), short-time objective intelligibility (STOI), and Failure rate (FR) \cite{delcroix2022listen}. FR measures the proportion of test samples with an SI-SDRi below 1 dB. Failures typically occur when the TSE system extracts the incorrect speaker or outputs the mixture.
\vspace{-2mm}
\subsection{Evaluation results following SUPERB's setup}
\label{sec:superb_exp}
\vspace{-1.5mm}
Table \ref{tab:superb} shows the performance of different SSL models using the SUPERB style downstream model describes in Section \ref{sec:SUPERB}.
We observe that SSL models significantly outperform the acoustic feature (STFT and FBANK) models across all three metrics: SI-SDRi, STOI, and PESQ in Table \ref{tab:superb}. This suggests the potential of utilizing SSL models for TSE tasks. 
WavLM models outperform Hubert and wav2vec versions, probably because the training style, including noise and interference speakers, provides more robust speech representations. Although WavLM Large achieves the best performance, we chose the more compact WavLM Base Plus in the remaining experiments.
Note that the best model achieves an SI-SDRi of 10.3 dB, which is significantly lower than TSE models trained from scratch, such as SpecEx+ \cite{ge2020spex} or TD-SpeakerBeam \cite{delcroix2020improving}, which attain an SI-SDRi of more than 13 dB.
%Among the SSL models, Wavlm Base Plus shows promising results. One potential reason is that, unlike other models that are pre-trained on single-source data, WavLM Base Plus incorporates data augmentation strategies involving noisy and multiple speakers. This data diversity enhances the model's robustness and generalization capability under mixture speech conditions. In addition, WavLM Large shows the highest performance. However, it is noted that a performance gap still exists between SSL-based models and current model (TD-SpeakerBeam) trained from scratch for TSE tasks.

%Moreover, we analyze the weights of both the Speaker Encoder and the Extractor, utilizing WavLM Base Plus as the backbone architecture. As shown in Fig \ref{fig:2}, for the Extractor module, which is responsible for estimating the STFT mask, the weights demonstrate a decreasing trend as the layer depth increases.

\begin{table}[t]
\caption{Performance comparison of various SSL models for target speech extraction tasks, following the SUPERB Challenge settings. }
\label{tab:superb}
\centering
\begin{tabular}{L{2.5cm}R{1.3cm}R{1.2cm}R{1.2cm}}
\toprule
\multirow{2}{*}{Upstream} & \multicolumn{3}{c}{Target Speech Extraction} \\ \cline{2-4} 
                          & SI-SDRi$\uparrow$         & STOI$\uparrow$          & PESQ$\uparrow$         \\ \midrule
STFT                      & 5.96            & 0.79         & 1.48        \\
FBANK                     & 5.18            & 0.78         & 1.41        \\ \midrule
HuBERT Base               & 9.18            & 0.86         & 1.77               \\
wav2vec 2.0 Base          & 9.15            & 0.86         & 1.76               \\
WavLM Base                & 9.69            & 0.87         & 1.95        \\
WavLM Base Plus           & 9.96            & 0.88         & 1.97        \\ 
WavLM Large               & 10.30           & 0.88         & 2.01        \\ \bottomrule
\end{tabular}%
    \vspace{-0.4cm}
\end{table}
% I will shorten the y-axis to compress the size of this fig
\vspace{-1.5mm}
\subsection{Evaluation Results on TD-SpeakerBeam setup}
\vspace{-1.5mm}
Table \ref{tab:mytable1} shows the extraction performance of different variants of TD-SpeakerBeam using or not using SSL features. We compare different versions of the SpkEnc, SSL features (from Transformer, CNN layers, or both), and fusion methods of the AIE module. The first row corresponds to the baseline TD-SpeakerBeam, which uses a TCN as SpkEnc. The second row replaces the SpkEnc with the SSL-MHFA introduced in Section \ref{sec:spkenc}. We observe that using SSL for SpkEnc degrades performance. One possible reason could be the difference in model architectures. The lightweight attentive pooling might be insufficient to effectively deal with the complex feature distributions captured by SSL-based models when paired with a TCN-based extractor that is randomly initialized. 

Next, we investigate augmenting the input of the extractor with different SSL features. Using a weighted sum of the transformer layers (followed by a deconvolution layer for upsampling) or the single CNN layer that matches the time resolution of the extractor fails to improve the results significantly. 
%We investigate the impact of utilizing intermediate layer representations from pre-trained CNN encoders and Transformer layers of Wavlm base plus for TSE tasks summarized in Table \ref{tab:mytable1}.
%From the table, it is observed that a simple replacement of the Speaker Encoder with an SSL-based model results in performance degradation. One possible reason could be the difference in model architectures. The lightweight attentive pooling might be insufficient to effectively deal with the complex feature distributions captured by SSL-based models when paired with a TCN-based based Extractor that is randomly initialized. 
To address this, we introduce the AIE module. %By directly cascading all CNN Encoder layer outputs, we achieve comparable performance (SI-SDRi 13.02 dB) with that of employing a weighted sum of Transformer layers. 
%In this approach, each layer's output is independently processed through a corresponding deconvolutional layer to achieve the same temporal resolution as the Encoder in the TSE. 
Notably, adopting upsampling modules to obtain hierarchical features, including U-Net and FPM, outperforms the baseline. Finally, merging the feature representations from both the Transformer and CNN Encoder layers offers additional performance improvements.
These results demonstrate that we can improve extraction performance with a pre-trained SSL model if we use a strategic integration method like our proposed AIE.
%Note that we confirmed that the gains were not due to the increased number of model parameters from AIE.
%provides significant performance gains. %Specifically, by initializing the top-most layer with a weighted sum of Transformer layers resulting in the highest SI-SDRi score.

\begin{table}[t]
\centering
\caption{Performance comparison for different approaches to exploit SSL models for TD-SpeakerBeam setup.} %Comparison of different methods utilizing intermediate layer representations of the pre-trained CNN encoder and Transformer layers for TSE tasks.}% The speaker encoder of baseline TD-speakerbeam consists of TCN.}
\label{tab:mytable1}
    \scalebox{0.9}{
\begin{tabular}{cllrr}
\toprule
\multirow{2}{*}{\begin{tabular}[c]{@{}c@{}}SpkEnc \\ \end{tabular}} & \multicolumn{2}{c}{AIE} & \multirow{2}{*}{SDR$\uparrow$} & \multirow{2}{*}{SI-SDRi$\uparrow$} \\
 & SSL Feature & Fusion Method &  &  \\ \midrule
TCN &  -& - & 13.69 & 13.03 \\ \midrule
\multirow{7}{*}{SSL-MHFA} & -& - & 12.91 & 12.13 \\ & Transformer & Weighted Sum & 13.84 & 13.18 \\ 
 & Single- CNN & - & 13.12 & 12.40 \\
 \cline{2-5} 
 & \multirow{2}{*}{Multi- CNN} %& Concat & 13.76 & 13.02 
   & Unet & 14.03 & 13.67 \\ 
 &  & FPM & 14.14 & 13.49 \\
 \cline{2-5} 
 & Multi- CNN & Unet & 14.38 & 13.80 \\
 & + Transformer & FPM & 14.65 & 14.01 \\
 \bottomrule
\end{tabular}}
    \vspace{-0.6cm}
\end{table}

%\begin{table}[t]
%\centering
%\caption{Detailed ablation study of different components of the HiT-CNN Encoder-Unet and SSL-based speaker encoder (SSL-MHFA). \textbf{MHFA} suggests multi-head factorized attentive pooling. }
%    \scalebox{0.85}{
%\label{tab:table2}
%\begin{tabular}{llc}
%\toprule
%Speaker Encoder & Adaptive Input Enhancer (multi-cnn)      & SI-SDRi$\uparrow$ \\ \midrule
%TCN             & -                            & 13.03   \\ \midrule
%TCN             & Unet         & 12.63   \\
%SSL-MHFA          & Unet {[}R{]} & 13.26   \\
%SSL-MHFA          & Unet         & 13.67   \\ 
%SSL-MHFA          & Unet+Transformer            & 13.80 \\
%SSL-MHFA          & FPM+Transformer           & %\textbf{14.01} \\

%\bottomrule
%\end{tabular}}
%    \vspace{-0.3cm}
%\end{table}

%\subsection{Ablation Studies on Architectures}
We also investigate the contributions of various components in our proposed architecture. Employing a UNet as an AIE with a traditional TCN-based SpkEnc yields an SI-SDRi of 12.63 dB, which is slightly lower than the baseline. This result demonstrates that while direct component replacement may not yield immediate benefits, strategic integration like AIE and SSL-MHFA provides significant performance gains. We also confirmed that a model with the same architecture as the AIE but randomly initialized performs worse than when using the pretrained SSL model (13.26 dB v.s. 13.67 dB).  %( With the same amount of CNN Encoder parameters, the performance of randomly initialized models underperforms than pre-trained initialization (13.26 dB v.s. 13.67 dB). 
This highlights the importance of the SSL pre-training in capturing robust features beneficial for TSE tasks.

\begin{table}[t]
\centering
\caption{SI-SDRi performance under different learning conditions. FR denotes Failure rate.}
\scalebox{0.85}{
\label{tab:learning-conditions}
\begin{tabular}{lccccc}
\toprule
\multicolumn{1}{c}{Fine-tuning} & SDR$\uparrow$ & SI-SDRi$\uparrow$ & STOI$\uparrow$ & PESQ$\uparrow$ & FR(\%)$\downarrow$ \\
\midrule
\multicolumn{1}{c}{\ding{55}}  & 14.65 & 14.01 & 0.91 & 2.38 & 3.9\\
\multicolumn{1}{c}{\ding{51}} & \textbf{15.26} & \textbf{14.65} & \textbf{0.93} & 2.45 & 3.0\\
\hline
TD-SpeakerBeam \cite{delcroix2020improving} & 13.69 & 13.03 & 0.90 & 2.12 & 4.8 \\ 
SpEx+ \cite{ge2020spex} & - & 13.41 & - & \textbf{2.93} & - \\ 
sDPCCN \cite{han2022dpccn} & - & 11.61 & - & - & -\\ 
\bottomrule
\end{tabular}
}
\vspace{-4mm}
\end{table}

\begin{figure}[]
    \centering
    \includegraphics[width=0.9\linewidth]{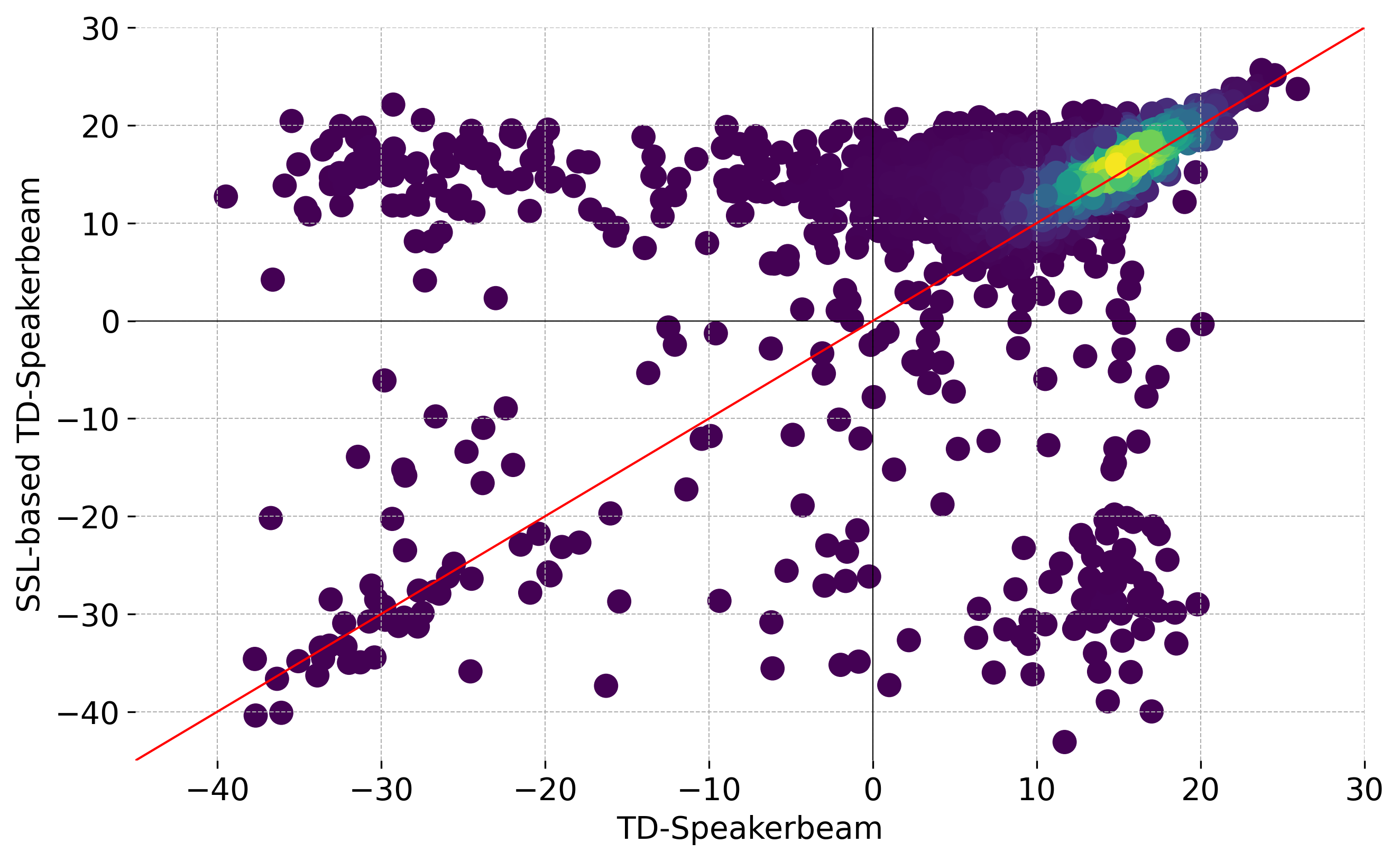}
        \vspace{-0.3cm}
    \caption{Comparison of SI-SDRi scores of test set samples using TD-SpeakerBeam (X-axis) against the best SSL-based model (Y-axis).}% Point color intensity represents the density of nearby samples.}
            \vspace{-0.4cm}
    \label{fig:enter-label}
\end{figure}

%\subsection{Impact of Learning Conditions on SSL-based Components}
We further investigate the performance of our proposed model after fine-tuning all parameters, including the SSL model. 
%under various learning conditions, specifically whether the SSL model is kept frozen or fine-tuned during training. 
The results are summarized in Table \ref{tab:learning-conditions}.  The best performance is achieved when both AIE and SpkEnc  are fine-tuned, reaching an SI-SDRi of 14.65. %Moreover, learning only the SpkEnc provides only little improvement. 
In side experiments, we confirmed that constraining the same SSL model for both AIE and SpkEnc results in about 0.1 dB SI-SDRi degradation but significantly reduces the number of model parameters. 
The proposed system significantly outperforms the baseline methods, including TD-SpeakerBeam, SpEx+, and sDPCCN, underscoring the advantages of incorporating SSL models into the TSE framework.
%^Moreover, constraining AIE and Speaker Encoder to share the same SSL backbone (\checkmark{*}) results in suboptimal performance compared to separate fine-tuning. This underperformance may be due to the specialized optimization requirements for each component, which can be better addressed by allowing independent backbones.

Finally, we analyze the performance improvement achieved by incorporating SSL into the TSE system. Figure \ref{fig:enter-label}  plots the SI-SDRi of TD-SpeakerBeam versus that of the proposed extension of TD-SpeakerBeam with a fine-tuned SSL model, where each dot represents the performance of one test sample. We observe that it not only improves the quality of the already well-extracted samples but also improves performance when TD-SpeakerBeam performs relatively poorly (SI-SDRi between -10 and 10dB). This translates into having a significantly lower FR value (3.0\% v.s. 4.8\%) as shown in Table \ref{tab:learning-conditions}, which indicates a more accurate identification of the target speaker. %, demonstrating the effectiveness of SSL-based speaker encoder.

%the performance improvement achieved by incorporating SSL into the TSE system. The red diagonal serves as a benchmark, denoting equivalent performance between the models. Most of the samples lie above the diagonal line, suggesting that the SSL-based model consistently outperforms TD-SpeakerBeam. In detail, the densely clustered points (in light colors) above the diagonal represent samples that have been separated well by TD-SpeakerBeam. Their prominence in the higher SI-SDRi region suggests that the SSL-based extractor further improves the quality of these already well-extracted samples, emphasizing its superior extraction capabilities. In addition, a lower FR value (3.0 v.s. 4.8) in Table \ref{tab:learning-conditions} indicates a more accurate identification of the target speaker, demonstrating the effectiveness of SSL-based speaker encoder.

\section{Conclusions}
In this work, we proposed using pre-trained SSL models for TSE. %introduced a novel approach to TSE that strategically incorporates representation obtained from a pre-trained SSL model.
%as both Speaker Encoder and AIE. 
We introduced a new downstream task, following SUPERB, as a benchmark for evaluating the performance of TSE models. Besides, we explored using SSL models with more powerful TSE systems. 
Our extensive experiments on Libri2mix demonstrate the importance of exploiting both CNN and Transformer layers of the SSL model and properly upsampling the representation. After fine-tuning SSL-based components, we improved significantly over existing systems trained from scratch. 
Future work will include exploring similar AIE modules for other SE tasks and reducing the number of parameters of the models.
%In future work, we plan to integrate the proposed architecture into the SSL pre-training stage and explore the potential of time-domain methods.

% Below is an example of how to insert images. Delete the ``\vspace'' line,
% uncomment the preceding line ``\centerline...'' and replace ``imageX.ps''
% with a suitable PostScript file name.
% -------------------------------------------------------------------------

%\vfill
%\pagebreak

\vfill\pagebreak

\footnotesize
\bibliographystyle{IEEEbib}
\bibliography{refs}

\end{document}